\documentclass[%
 reprint,
 amsmath,amssymb,
 aps,
floatfix,
]{revtex4-2}

\usepackage{graphicx}
\usepackage{dcolumn}
\usepackage{bm}
\usepackage[flushleft]{threeparttable}

\begin{document}

\preprint{APS/123-QED}

\title{Anti-Neutrino Flux from the EdF Hartlepool Nuclear Power Plant}

\author{Sandra Bogetic}%
 \thanks{co-author}%
 \email{sbogetic@utk.edu}
 \collaboration{ WATCHMAN Collaboration}
\affiliation{
 University of Tennessee, Knoxville, Tennessee 37996, USA 
}%

\author{Robert Mills}
 \thanks{co-author}%
 
 \email{robert.w.mills@uknnl.com}
 \collaboration{ WATCHMAN Collaboration}
\affiliation{%
 UK National Nuclear Laboratory, Central Laboratory, Sellafield, Seascale, Cumbria, CA20 1PG, UK}


\author{Adam Bernstein}
\affiliation{
Lawrence Livermore National Laboratory, Livermore, California 94550, USA
}%
\collaboration{ WATCHMAN Collaboration}

\author{Jonathon Coleman}
\affiliation{%
 University of Liverpool, Liverpool, L69 7ZE, UK
}%
\collaboration{ WATCHMAN Collaboration}

\author{Alex Morgan}
\affiliation{%
University of Liverpool, Liverpool, L69 7ZE, UK
}%

\author{Andrew Petts}
\affiliation{%
 EDF Energy, Barnett Way, Barnwood, Gloucester GL4 3RS, UK\\
}%

\date{\today}

\begin{abstract}
In this article, we present the first detailed simulation of the antineutrino emissions from an Advanced Gas-cooled Reactor (AGR) core, benchmarked with input data from the UK Hartlepool reactors. An accurate description of the evolution of the antineutrino spectrum of reactor cores is needed to assess the performance of antineutrino-based monitoring concepts for nonproliferation, including estimations of the sensitivity of the antineutrino rate and spectrum to fuel content and reactor thermal power. The antineutrino spectral variation we present, while specific to AGRs, helps provide insight into the likely behavior of other  reactor designs that use a similar batch refueling approach, such as those used in RBMK, CANDU and other reactors.  

\end{abstract}

\maketitle


\section{\label{sec:level1}Introduction}





Nuclear reactors have been successfully used as a source of electron antineutrinos in numerous antineutrino physics experiments~\cite{Eguchi:2002dm,Abe:2011fz} and have been proposed for use in safeguards applications~\cite{Borovoi:1978,Bernstein:2008tj,An:2017osx}. Reactors have played a central role in characterizing antineutrino oscillations, due to their high intensity and reasonably well predicted antineutrino emission spectrum. The spectral prediction is essential for some monitoring applications and fundamental studies of antineutrino properties. For example, the spectral evolution can be used to estimate reactor characteristics such as fuel burnup and  time-averaged power~\cite{Borovoi:1978}, as well as facilitate the estimation of fundamental properties of the antineutrino, such as mixing angles~\cite{Dwyer_2015}. 

Antineutrinos arise from beta-decays of fission product daughters, with an average number of decays (and thus antineutrinos) of about three per daughter and six per fission. The total number of antineutrinos per fission varies only modestly with fissile isotope. However, the population of daughters varies considerably among these isotopes, resulting in substantial differences in the emitted antineutrino energy spectrum for each. To predict the spectrum, the number of fissions arising from each parent isotope - which varies in time over the course of the reactor cycle - must be calculated. In this article, we  perform this calculation through the  use of an assembly-level simulation of the core, which is used to track the evolution of fission rates from each isotope throughout the cycle. Antineutrino spectra at each instant in the cycle can then be derived by convolving the fission rates with the tabulated number of antineutrinos per unit of energy per fission~\cite{Huber:2011wv}.   

Other authors have made similar estimates for Pressurized Water Reactors fueled with low-enriched uranium~\cite{PhysRevD.86.012001},  mixed oxide assemblies~\cite{MOXspec},   CANDU reactors~\cite{MATTHEWS201367}, and thorium-fueled reactors~\cite{akindele2016thorspec}. The prediction presented in this article is the first such estimate for an Advanced Gas-cooled Reactor (AGR).  A key finding is that the relatively short ($\sim$ \!4 month) refueling intervals and relatively small ($\sim 6\%$ of the core per outage) fuel replacement fraction contribute to a smaller change in the antineutrino flux and spectrum from beginning to end of cycle compared to other reactor types. 

\section{Hartlepool Advanced Gas-Cooled Reactor}

In United Kingdom, Germany, and United States Gas-cooled reactors have been in operation for many years. Specifically, in the United Kingdom, nuclear electricity has mostly been generated by CO$_2$ cooled MAGNOX and AGRs since the 1980s. The AGR considered for the antineutrino prediction study in this paper is the Harlepool Nuclear Power Plant. The choice of the Hartlepool reactor has been driven by the collaboration effort with the reactor engineers to get reactor operational data for the antineutrino study. This allowed for a high fidelity calculation of the reactor fractional fission rates for the most important fissioning nuclides in the reactor and to further the sensitivity study of the antineutrino flux on the level of fidelity of the reactor simulations.

\subsection{Reactor Description and Operation}

Hartlepool Power Station is located on the North-East coast of England and has been safely producing low-carbon electricity since 1983. The power station operates two AGRs. The AGR design at Hartlepool Power Station consists of a graphite-moderated reactor core and uses pressurised CO$_{2}$ as the primary coolant. The mean diameter of the active core is approximately 9.3\,m and 8.2\,m high, with total core diameter being approximately 11.9\,m and 12.7\,m high. The reactor core consists of columns of circular cross-sectional graphite bricks with interstitial square cross-sectional graphite bricks. There are 324 on-lattice fuel channels formed by bores in the larger circular bricks. Each bore is 0.27\,m in diameter and is pitched at approximately 0.47\,m. 81 control rod channels are in a one-in-four array in the smaller square brick columns. Within each of the 324 fuel channels are 8 stacked fuel elements, each containing 36 clustered fuel pins arranged in concentric rings of 18, 12 and 6 pins within a graphite sleeve. The stainless-steel fuel pins are approximately 1\,m in length with a diameter of 14.48\,mm and contain stacked ceramic UO$_{2}$ pellets of either 3.2\% or 3.78\% $^{235}$U. The total core inventory of uranium is approximately 130 tonnes. The primary coolant is driven around the core by eight gas circulators which each have a constant-speed motor running at 3000\,rpm, resulting in a total gas mass flow of 3600\,kgs$^{-1}$. In normal, full-power operation, the primary coolant operates at 39 bar with temperatures at the bottom of the active core around 543\,K and at the top of the active core 923\,K. Heat is deposited into eight steam generators by the pressurized CO$_{2}$. Demineralised water acts as the secondary coolant and is fed to each steam generator in a closed-loop system at a rate of 60\,kgs$^{-1}$. The primary and secondary loops are separated within the steam generator to ensure no contamination is spread to the steam-side equipment from the active primary coolant. The steam generators operate at approximately 139 bar and produce steam at 843\,K resulting in an output of around 1560\,MW$_{th}$ and 600 MW$_{e}$ for each reactor.
\newline
Fuel typically remains in the core for around eight years, with an average discharge irradiation of 32\,GWd/T$_{e}$. On-load average fuel irradiation is around 15.5\,GWd/T$_{e}$. The reactor typically remains at full power for 20–22 weeks before undergoing a controlled reactor shut down for off-load, depressurized batch refueling where around 20 of the highest burnup fuel assemblies are replaced. The reactor remains shut down during these periods for around 10-14 days, after which it returns to full-power operation over a period of a 2-3 days. Outages across the two reactors are staggered to avoid overlap of shutdown periods. This results in five shutdown periods per year across the two reactors.

\subsection{PANTHER Simulation of the Full Core}

Full operational data for both Hartlepool reactors were obtained including power history, fuel element irradiation (MWd/t$_e$), power rating (MW/t$_e$), initial enrichment and loading dates for the whole core over a 12 month operational period. Reactor power was calculated using the thermal hydraulics code HEYPEX. The HEYPEX code calculates reactor power based on plant measurements of coolant mass flow and inlet and outlet reactor core temperatures. The thermal-neutronics code PANTHER was used to calculate individual fuel element powers and irradiations as part of the Core Follow Regular Assessment Route. The Regular Assessment Route is executed approximately weekly and provides a steady-state assessment against various reactor compliance limits. A fixed, 3D reactor model describing reactor geometry, materials and basic nuclear data is imported into the calculation route. The reactor model consists of 1 meshpoint per reactor channel, providing 324 radial points. Axially, the reactor model consist of 8 meshpoints, 1 per fuel element.  Measured plant parameters such as control rod positions, reactor thermal power, fuel channel inlet and outlet temperatures and coolant mass flows are imported along with the 3D reactor state from the previous Regular Assessment, which is used as a calculation starting point. Individual element powers, irradiations and decay heat source terms are then calculated using the thermal-neutronic model to solve a steady-state diffusion equation approximation.


\subsection{Prediction of Fission Rates Using FISPIN}

The Schreckenbach measurements reported in 1981, 1982 and 1989 \cite{Schreckenbach:1994} showed little change in the emitted beta spectra for the neutron irradiation of pure actinide samples after 20 to 30 hours. It being assumes that the beta emission is dominated by short-lived fission products that for the case of a constant fission rate of a specific nuclide will quickly produce an equilibrium concentration of the resultant fission products and thus a time-independent spectra thereafter. More recently, Huber \cite{Huber:2011wv} used these measured spectra and the observation to assume that the neutrino emission of reactors can be represented by an anti-neutrino spectra for each nuclide being fissioned. Thus if you know the fission rates for each principle fissionable nuclides you can use these to weight the time independent anti-neutrino spectra per fission  to produce an anti-neutrino emission spectra from a reactor.

In previous work Mills et al \cite{MILLS20202130} used the UK reactor modeling code WIMS and the UK spent fuel inventory code FISPIN to model the change in fission rates in nuclear fuel with burnup for typical initial enrichment, reactor power and burnup values for Pressurized Water Reactor (PWR), Boiling Water Reactor (BWR) and AGR reactors in a fuel assembly. These results being published as a database in Mendeley Data \cite{Mills2020md}.

For a specific reactor type it is therefore possible to interpolate the fission rates in a reactor assembly based upon initial enrichment, burnup and power. From the fission fractions the average energy release per fission can then be determined and thus the total number of fissions estimated for a given power.  If these data are available for all the fuel assemblies in a reactor, the total anti-neutrino source can be determined using the Huber anti-neutrino spectra for each fissioning nuclide per fission. 

\subsection{Antineutrino Rate}

Antineutrinos from the Hartlepool reactor are here detected via the inverse beta decay (IBD) interaction \cite{Huber:2015ouo,Huber:2011wv}. In this process, antineutrinos interact with quasi-free protons in the water, producing a positron- neutron pair in the final state:
\begin{equation}
\overline{\rm \nu_e} + p \rightarrow e + n
\end{equation}

The positron is detected as a prompt signal through the Cherenkov light emitted when the particle velocity exceeds the speed of light in the water. This is a coupled threshold reaction in which the antineutrino energy must exceed 1.8 MeV to generate an IBD reaction, and the resulting positron kinetic energy must exceed $\sim$253 keV in water to generate Cherenkov light. The neutron produced through IBD will elastically scatter off hydrogen in the detector until thermalization, after which it can capture on either a gadolinium or hydrogen nucleus.  Following gamma-ray production from neutron captures, Cherenkov light is emitted through the Compton scattering on electrons. Due to the threshold required for Cherenkov emission, not all of the scattered electron from gamma-rays released from neutron capture on Gd will contribute to the signal. From there the individual events were sampled based on the reactor thermal power, and standoff from the detector, L. The expected antineutrino flux N (E $_{\overline{\nu}}$ ) detected in the detector is given by:

\begin{equation}
\centering
N(E_{\overline{\rm \nu_e}},L)=\dfrac{n_p T}{4\pi L^2} \sum_{l}N_l ^f \phi_l (E_{\overline{\nu}})\sigma(E_{\overline{\nu}})P_{ee}(E_{\overline{\nu}},L)
\end{equation}

Here, n$_p$ refers to the number of quasi-free protons and T is the counting time of the experiment. The electron antineutrino survival probability due to oscillations and the inverse beta decay cross section is given by P$_{ee}$ and $\sigma$ respectively \cite{Huber:2011wv}. The individual contributions of fissile isotopes $l$ is represented by the fissile fraction for the specific isotope N$_l ^f$ and the unique spectra for that isotope $\phi_l$, which is assumed to follow the approximations taken from Reference \cite{Huber:2011wv}.

Neglecting contributions from the backgrounds, \\ the measured antineutrino signal will be governed by the position, burnup, and power of Hartlepool NPP. The analysis assumes the following energy per fission contribution: $^{235}$U (201.7 MeV/Fission), $^{238}$U (205.0 MeV/Fission), $^{239}$Pu (210.0 MeV/Fission), and $^{241}$Pu (212.9 MeV/Fission) used by FISPIN.

\section{Hartlepool ANTI-NEUTRINO emission estimation}

The reactor information supplied by EDF Energy including the burnup (GWd/t), power (MW/t), initial enrichment ($^{235}$U/U weight percentage) for each of the 2592 assemblies in both reactors at a selection of times during the 2020 calendar year as well as the total thermal reactor power (MW) during this period. Thus the anti-neutrino emission from each assembly can be determined and combined to produce an anti-neutrino emission spectra for each reactor. Noting that you must calculate the emission for each assembly and combine these before recalculating the fission fractions from the total to give core average values as the energy per fission will vary during burnup differently in each assembly.


%

\begin{figure}[h]
\includegraphics[scale=0.33]{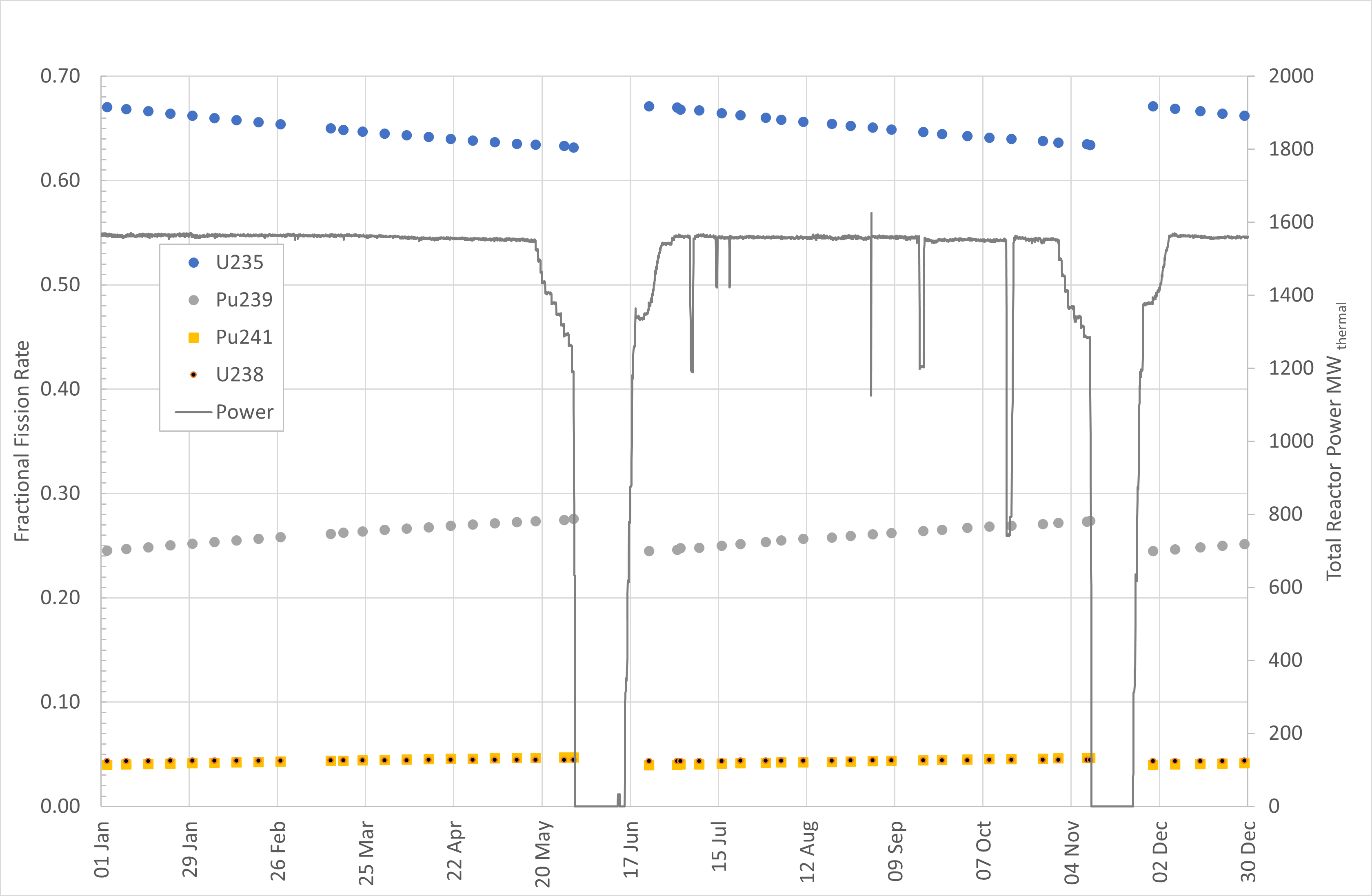}
\caption{\label{fig:wide}Fractional Fission Rates for the Hartlepool reactor unit 1 in 2020.}
\label{fig:FFR1}
\end{figure}

\begin{figure}[h]
\includegraphics[scale=0.33]{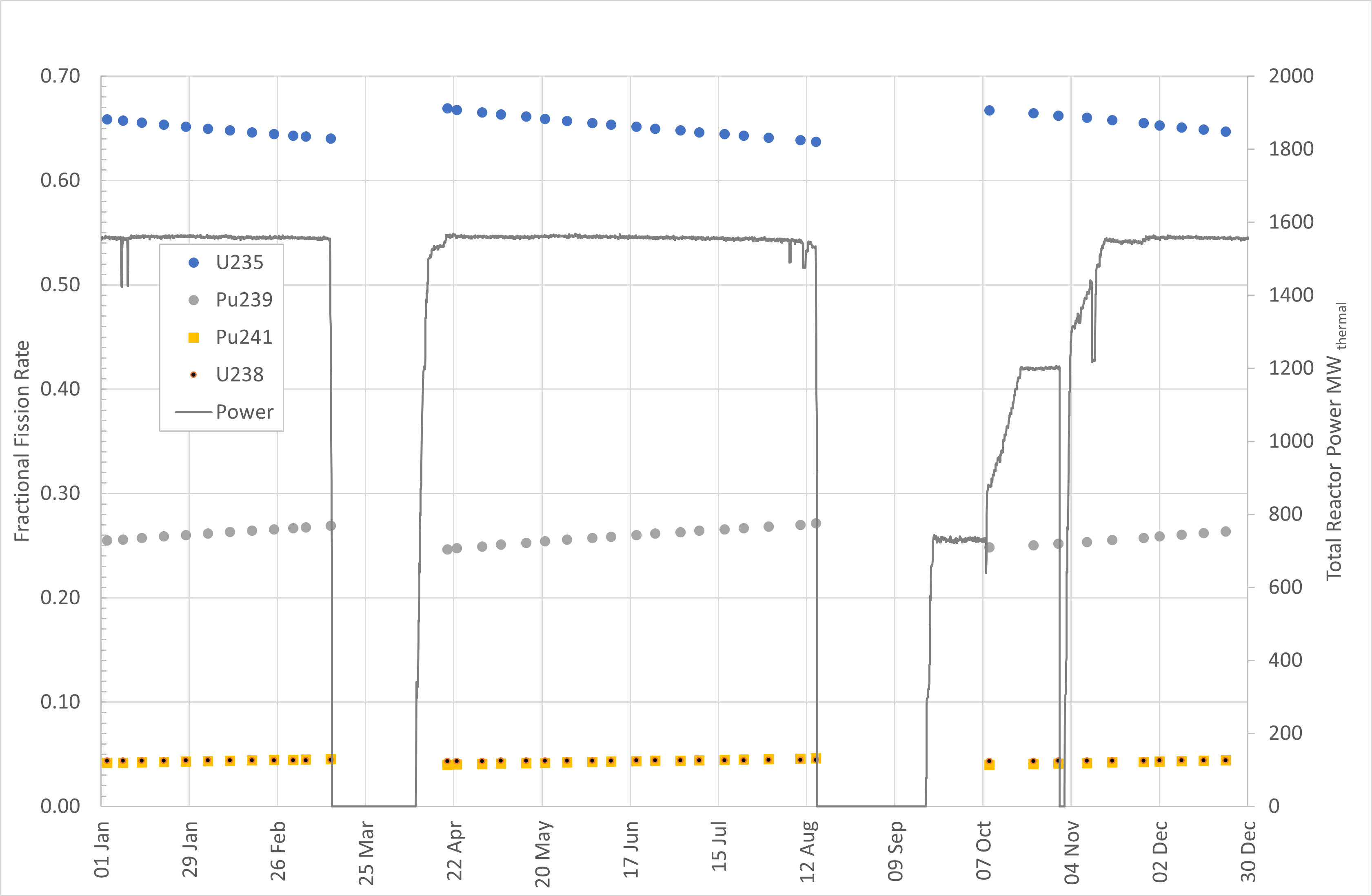}
\caption{\label{fig:epsart} Fractional Fission Rates for the Hartlepool reactor unit 2 in 2020.}
\label{fig:FFR2}
\end{figure}

The reported reactor power over calendar year 2020 and the estimated variation of total core fission fractions are shown in figures \ref{fig:FFR1} and \ref{fig:FFR2}. It should be noted that approximately 12\% of the assemblies in the core are changed during shutdowns approximately every 6 months, so that the fractional fission rates in the locations of replaced fuel are reset to the values of zero burnup at these times.  The frequent refueling of a small fraction of the core results in the fractional fission rates varying less than reactors that change a greater fraction of the core yearly or less frequently.

\subsection{Anti-neutrino Rate for Hartlepool}

\begin{figure}[h]
\centering
\includegraphics[scale=0.205]{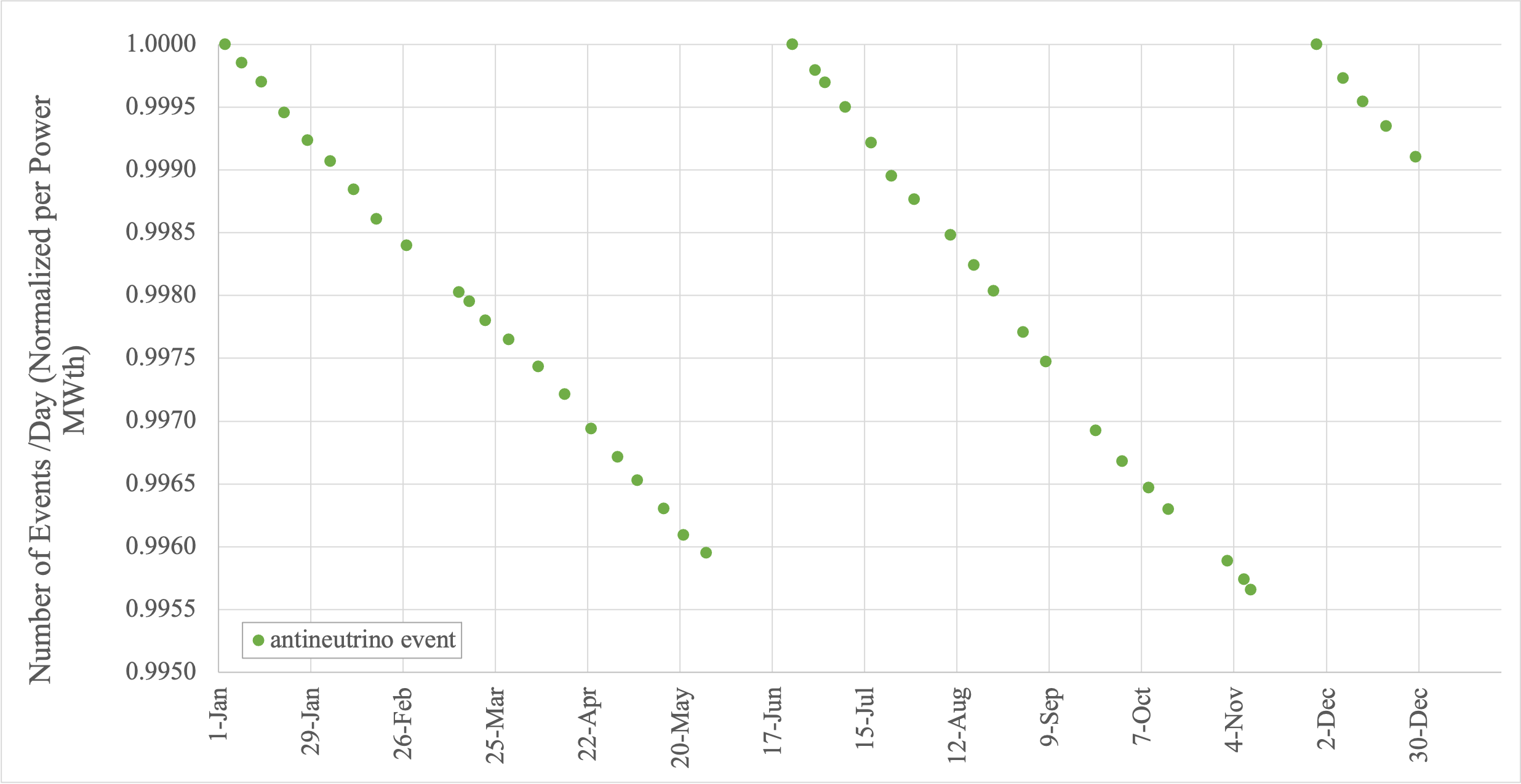}
\caption{Total antineutrino flux per reactor per fuel cycle length (days) for the Hartlepool AGR unit 1 in 2020}
\label{fig:flux1}
\end{figure}

\begin{figure}[h]
\centering
\includegraphics[scale=0.205]{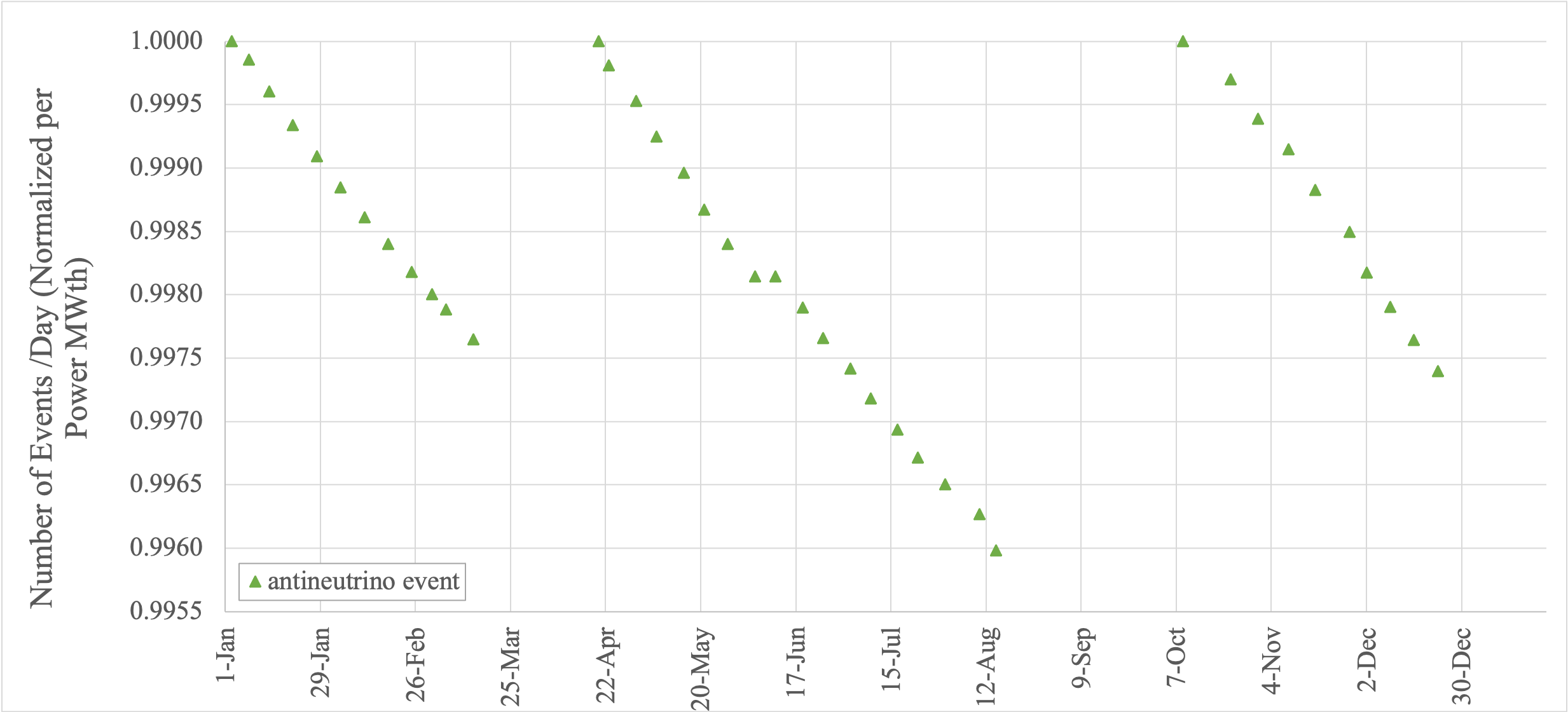}
\caption{Total antineutrino flux per reactor per fuel cycle length (days) for the Hartlepool AGR unit 2 in 2020}
\label{fig:flux2}
\end{figure}

The distance for which the neutrino rate is simulated is 26 km, which also represents the distance between the Hartlepool AGR and the STFC Boulby Underground Laboratory, whose low background at 1,100m depth could simplify antineutrino measurements at distance. Figure \ref{fig:flux1} and \ref{fig:flux2} shows the antineutrino rates for core 1 and for core 2.

\begin{figure} [h]
\centering
\includegraphics[scale=1.25]{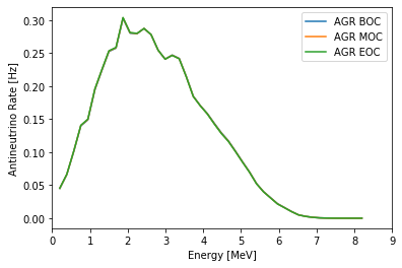}
\caption{Hartlepool AGR Spectrum Result of Anti-neutrino  Rates over Energy [MeV] along three points in the Reactor Fuel cycle (days: 1, 80, 160)}
\label{fig:spectra1}
\end{figure}

As shown in Figure \ref{fig:flux1} and \ref{fig:flux2} the antineutrino flux decreases, over the reactor cycle, with burnup, in accordance to LEU reactors. This is due to increased production of and reliance on $^{239}$Pu which emits fewer antineutrinos per fission above the inverse-beta detection threshold. On the other hand, the detected antineutrino rate is seen not to decrease significantly, due to the short reactor cycle and lower discharge burnup.  Figure \ref{fig:spectra1} presents the difference in antineutrino spectra shape from beginning (BOC) at day 1, middle (MOC) at day 80 and end of cycle (EOC) at day 160 of operation. As expected there are no evident difference in the shape or intensity of neutrinos spectra from BOC to EOC in either of the two cores. As explained above, only the combined effects of power and burnup are constrained by the measured antineutrino rate evolution. 

\section{Discussion: Lessons learned for Detection of reactor neutrinos}
\subsection{Rate from Full Core Analysis vs Simplified Model}

The possibility to have full core operational data for a year of operation gave a unique possibility in this research, to calculate high fidelity antineutrino rates for a full cycle evolution and to determine the sensitivity of the antineutrino evolution to the reactor full core operation calculation.  In practice, the anti-neutrino emission from a reactor depends on the 3D distribution of fission rates throughout the core. These parameters themselves depend on the fuel composition and neutron flux within the core. This required a detailed burnup analysis of the reactor including knowledge of the fuel placed within the core with its design, including initial composition, and operations parameters, as well as temperatures of the fuel and coolant, and the reactor thermal power. In practice, this is information that is only known to those modeling the reactor to support operation and seldomly distributed to others. 

\begin{figure} [h]
\centering
\includegraphics[scale=0.33]{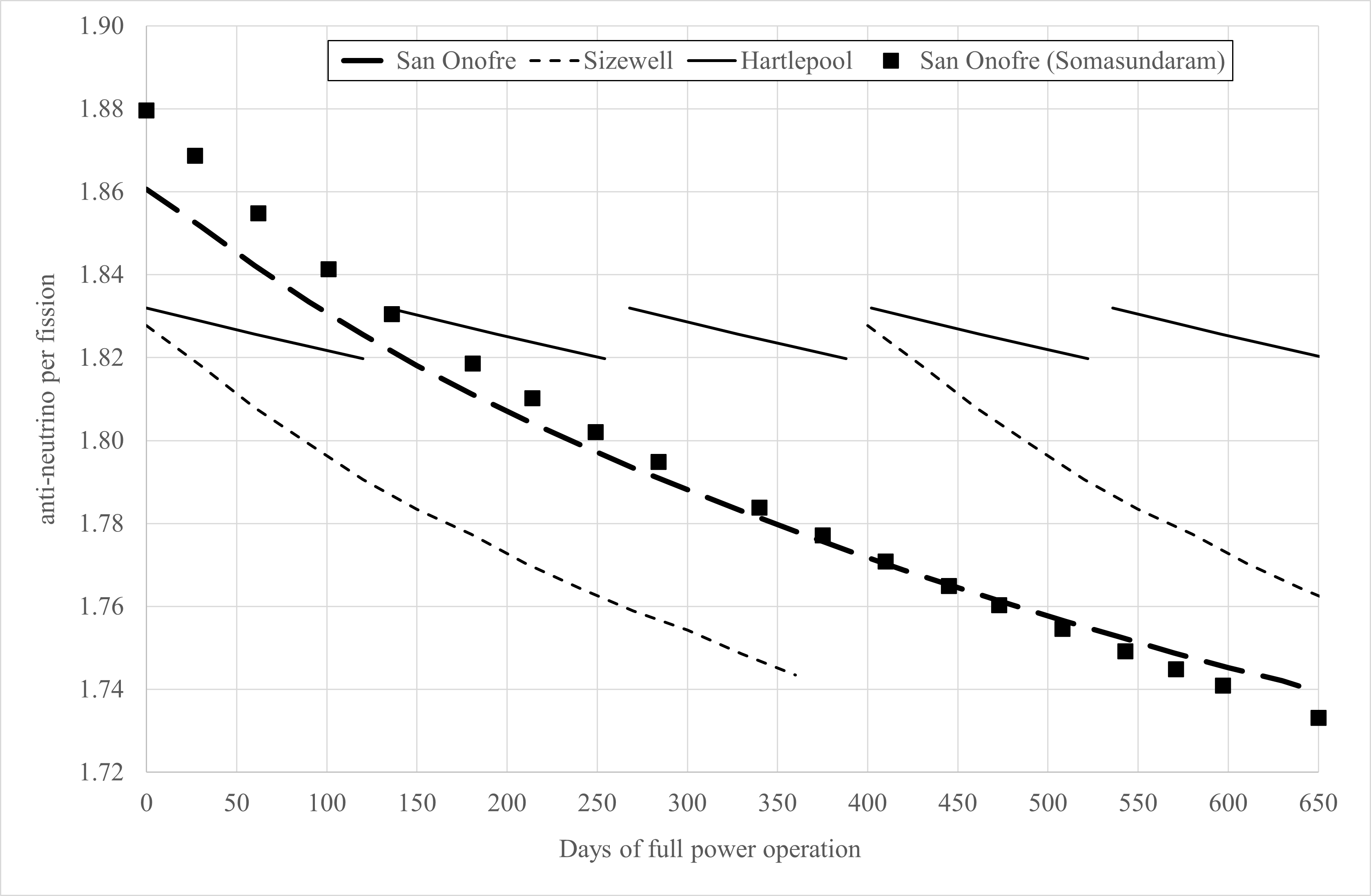}
\caption{Variation of the detectable anti-neutrinos per fission from the Hartlepool, Sizewell and San Onofre reactors calculated using the simplified method convoluted with the Huber~\cite{Huber:2011wv} number of neutrinos emitted per fission over the IBD threshold of 1.8 MeV. Convolution of the detailed whole core model results using the CASMO/SIMULATE package~\cite{Soma2012} and~\cite{Soma2011} are shown by the square markers.}
\label{fig:NuCompare}
\end{figure}

For this study, we are able to perform a comparison using a simplified model of a Hartlepool AGR, the Sizewell B PWR and the San Onofre PWR operations. These PWRs being chosen as they use different refueling strategies, but similar fuels. The approximation used is that the core consists of a number of fuel batches that when inserted have zero burnup, but then accumulate burnup according to the average power of the reactor. So if the core consists of N batches of fuel, each batch would have a different burnup depending on the number of cycles it has been in the core. The fractional fission rates can be determined at the start-up following a fuel replacement for each batch and then during the period until the reactor is shutdown for the next fuel replacement. Using the available data in Table \ref{tab:ReactInfo}  the changes of the  antineutrino emission was calculated and shown in Figure \ref{fig:NuCompare}.  The AGR shows less variation with its more frequent refueling, with the yearly and two yearly PWR showing a much more pronounce variation. It is noted that any independent calculation previously reported using the CASMO/SIMULATE package gives very similar results to this simplified method using the WIMS/FISPIN fractional fission rates based upon different nuclear data libraries. 

\begin{table*}
\centering
\caption{Reactor Information from the World Nuclear Industry Handbook 2003 \cite{tarlton2003world}} 
  \begin{threeparttable}[t]
\begin{tabular}{ |p{7.8cm}||p{2cm}|p{1.8cm}|p{1.9cm}|p{1.5cm}|  }
 \hline
 Quantity & San Onofre 2 & Sizewell &  Hartlepool 1 & Units \\
 \hline\hline
 Reactor type &PWR&PWR&AGR&-\\\hline
Enrichment&3.97*&3.1&3.1&Weight\% $^{235}$U/U \\\hline
Mass of initial uranium in the core&89.5&88.6&110&tons\\\hline
Thermal reactor power&3390&3411&1550&MW\\\hline
Average rating &37.9&38.5&14.09&MW/t\\
(Thermal power divided by initial uranium mass)& & & & \\\hline
Length of cycle&24&12&4&Months\\\hline
Typical shutdown between cycles&60&40&14&Days\\\hline
Percentage of fuel mass changed at refueling&49.5&33&7.3&\%\\\hline
The typical number of cycles fuel is in the core, N&2&3&14&-\\
(Approximated to the nearest integer) & & & & \\\hline
Average burnup of spent fuel discharged&33&33&24&GWd/t\\\hline
\end{tabular}
\begin{tablenotes}
\item [*] *Value not available for San Onofre 2, value from sister station San Onofre 3 used.
\end{tablenotes}
\end{threeparttable}
\label{tab:ReactInfo}
\end{table*} 

\subsection{Anti-Neutrino Flux Comparison AGR vs. PWR }

\begin{figure}[ht]
\centering
\includegraphics[scale=0.26]{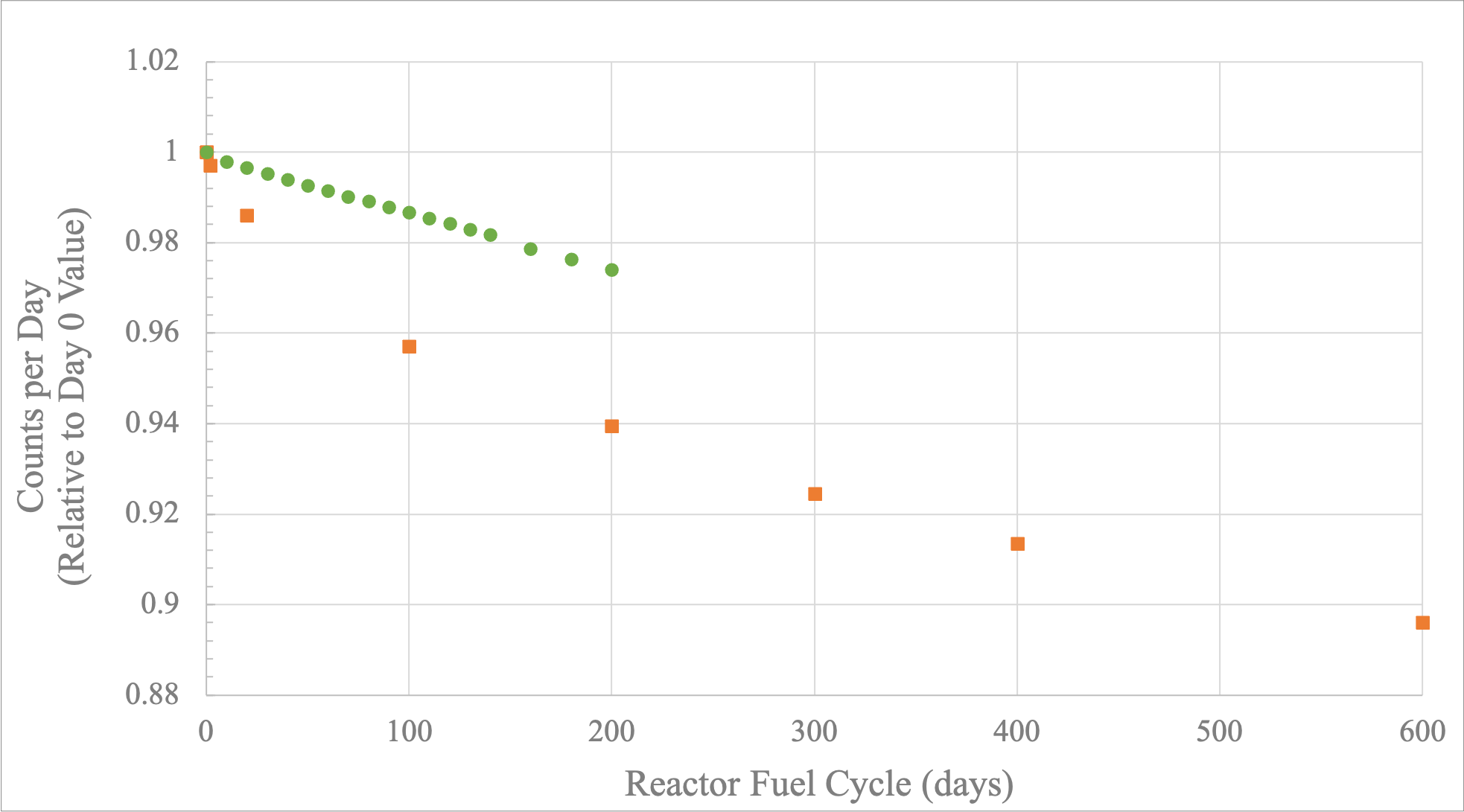}
\caption{Antineutrino flux rate for the Hartlepool reactor unit 2 and the SONG NPP}
\label{fig:AGR_SONG}
\end{figure}

\begin{figure}[ht]
\centering
\includegraphics[scale=1.22]{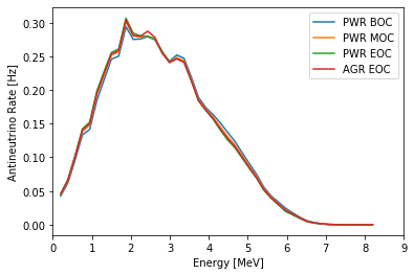}
\caption{Antineutrino Spectra for the SONG reactor unit}
\label{fig:spectra_SONG1}
\end{figure}

It is of interest for the sake of studying the sensitivity of the antineutrino rate for various type of nuclear power plants to compare the flux from an AGR to a Light Water Reactor commonly used around the world. The reactor chosen for the comparison is the SONGS NPP and the resulting antineutrino flux rate from the SONG detector experiment. The SONGS1 detector [13] was operated at the San Onofre Nuclear Generating Station (SONGS) between 2003 and 2006. The active volume comprised 0:64 tons of Gd doped liquid scintillator contained in stainless steel cells. The detector was located in the tendon gallery of one of the two PWRs at SONGS, about 25 m from the reactor core. As most operating PWR, SONGS have low-enriched uranium (LEU) cores with a
mixture of fissions: $^{235}$U ($\sim$55\%), $^{239}$Pu ($\sim$30\%), $^{238}$U ($\sim$10\%), $^{241}$Pu ($\sim$5\%), large power: $\sim$3 GWth and long fuel cycle length before refueling $\sim$600 days. For the sake of comparison, the SONGS PWR Reactor is considered at the same 26 km distance as Hartlepool/Boulby. As well the comparison seen in Figure \ref{fig:AGR_SONG} and \ref{fig:spectra_SONG1} the difference of flux along the cycle has been normalized to the initial flux for both cases (AGR vs. PWR). 

Under standard operation of both the PWR and AGR, the antineutrino flux rate  from primarily fissioning $^{235}$U to $^{239}$Pu can be tracked, and is in line with reactor observation. Changes in the anti-neutrino flux throughout an actual AGR fuel cycle and in a whole year is within 1\% in comparison to PWR changes of almost 10\%.  Also changes in the FFR of $^{235}$U and $^{239}$Pu are not significant in AGR due to the smaller fraction of the replaced fuel more regularly. The shorter fuel cycle of AGR respectively to the PWR: 210 vs 600 days are a key factor for a low antineutrino rate change from BOC to EOC, as the lower power $\sim$1500 MWth for AGR vs $\sim$3 GWth for the SONG PWR.

\section{Future of Reactor Monitoring}

Though only operated in the UK, the AGR, with its batch refueling regime is similar in operation to other reactor designs which employ on-load, small-batch refueling such as the RBMK, CANDU, MAGNOX, UNGG and BN series fast breeder reactors. One could therefore expect a similar antineutrino profile from such reactor designs. It is also worth noting that more exotic reactor designs such as pebble bed and some Generation IV reactors also employ a similar on-load refueling regime. 
\newline
It is known that plutonium production favors low-burnup, early discharge fuel and that the RBMK and MAGNOX reactor designs were developed in part for their ability to produced significant quantities of weapons-grade plutonium. More modern, conventional reactor designs favor steady operation over extended periods to maximize electricity generation, or capability factor and fuel utilization. Changes to the observed antineutrino profile could potentially be used to infer changes in refueling strategy, with a flatter profile inferring more frequent refueling and discharging of low-burn up fissile material. This has the potential to act as a verification tool to ensure operators of nuclear facilities are declaring accurate fuel loading and discharge information for nuclear safeguarding.
\newline
The information can be inferred from the measured antineutrino flux, in order to make reliable conclusions about reactor operations and fuel loadings, burnup and reactor power must be decoupled. Reported reactor power could be used for this, though is susceptible to falsification. Radioisotope emissions, and isotopic ratios could potentially be used to infer reactor power as certain key short-lived isotopes of xenon and krypton pro-rata with reactor load and are approximately independent of burnup. This information could therefore help interpret any observed changes in antineutrino profile. The XENAH project \cite{XENAH} currently in operation at Hartlepool Power Station is measuring isotopic emissions both at-source and remotely. Data from this project could provide valuable contextual information to either calculated antineutrino profiles reported in this work, or in future observed antineutrino profiles as measured by a detector deployed close to source.

\section{Summary and Conclusion}

In this article, we have presented the first detailed simulation of the antineutrino emissions from an AGR core, benchmarked with input data from the UK Hartlepool reactor.  Hartlepool operational data was provided to us for a full 12-month refueling and outage cycle, and this information was used as an input to the thermal hydraulics code HEYPEX and thermal neutronics code PANTHER, to calculate overall reactor power, and assembly-level power and burnup respectively. Emitted antineutrino spectra per fission and isotope are based on the Huber-Muller parametrization. These isotope-specific spectra are weighted by the number of fissions for each isotope, then summed to obtain an aggregate spectrum. The summed antineutrino spectrum evolves throughout the cycle due to the changing quantities of fissile isotopes in the core.    

One of our main findings is that the relatively short ($\sim$4 month) refueling intervals and relatively small ($\sim$6$\%$  of the core per outage) fuel replacement fraction in AGRs together lead to of a smaller change in the antineutrino flux and spectrum from beginning to end of cycle compared to other reactor types.

We provide this description of the evolution of the AGR antineutrino spectrum  to permit reliable assessments of the performance of antineutrino-based monitoring concepts for nonproliferation-oriented monitoring of AGRs, including estimations of the sensitivity of the antineutrino rate and spectrum to fuel content and reactor thermal power. The antineutrino spectral variation we present, while specific to AGRs, also helps provide insight into the likely behavior of other reactor designs that use a similar batch refueling approach, such as RBMK, CANDU and other reactors.

\begin{acknowledgments}

This work was performed under the auspices of the U.S. Department of Energy by Lawrence Livermore National Laboratory under Contract DE-AC52-07NA27344. Dr. Mills acknowledges funding from his laboratory's internal research and development program. The University of Liverpool authors acknowledge support by the STFC, and UK National Nuclear Laboratory for co-funding an EPSRC Next Generation Nuclear studentship. The authors would like to thank Sebin John and Daniel Gura of EDF Energy for their assistance in the production of the Hartlepool reactor data set.

\end{acknowledgments}

\appendix



\bibliography{refsAGR}

\end{document}